\documentclass[pre,onecolumn,11pt]{revtex4-1}
\usepackage{mathrsfs}
\usepackage{amsmath,amssymb}
\usepackage{bm}
\usepackage[dvipdfmx]{color}
\usepackage[dvipdfmx]{graphicx}
\usepackage{url}

\begin{document}
\title{Numerical method toward fast and accurate calculation of dilute quantum gas using Uehling-Uhlenbeck model equation}
\author{Ryosuke Yano}
\affiliation{Medical Academia, 1-38-12 Takata, Toshima-ku, Tokyo 171-0033, Japan}
\begin{abstract}
\textcolor{black}{The numerical method toward the fast and accurate calculation of the dilute quantum gas is studied by proposing the Uehing-Uhlenbeck (U-U) model equation. In particular, the direct simulation Monte Carlo (DSMC) method is used to solve the U-U model equation. The DSMC analysis of the U-U model equation surely enables us to obtain the accurate thermalization using a small number of sample particles and calculate the dilute quantum gas dynamics in practical time. Finally, the availability of the DSMC analysis of the U-U model equation toward the fast and accurate calculation of the dilute quantum gas is confirmed by calculating the viscosity coefficient of the Bose gas on the basis of Green-Kubo expression or shock layer of the dilute Bose gas around a circular cylinder.}
\end{abstract}

\maketitle

\section{Introduction}
The physical interests of the quantum gas increase in accordance with the developments of the study of the cold Bose gas in Bose-Einstein condensation \cite{Dalfovo}, cold Fermi gas \cite{Cao}, or quark-gluon plasma \cite{Hatsuda}. In particular, the Uehling-Uhlenbeck (U-U) equation \cite{Uehling} is significant for understanding of the characteristics of the dilute and \textcolor{black}{non-condensed quantum gas. As a result, we never mention to the Gross-Pitaevskii equation and its related equations for the condensate \cite{Blakie} \cite{Sinatra} \cite{Cockburn} in this paper. Thus, we focus on the dynamics of the dilute and non-condensed quantum gas, which is far from equilibrium state, exclusively.} As seen in previous studies on the U-U equation \cite{Arkeryd} \cite{Nikuni2}, the consideration of the characteristics of the U-U equation in the framework of the kinetic theory is somewhat depressing owing to the markedly complex collisional term of the U-U equation. Therefore, the quantum kinetic equations, which simplify the collisional term in the U-U equation, were proposed such as the quantum Fokker-Planck equation (FPE) \cite{Kaniadakis} \cite{Yano1} or quantum Bhatnagar-Gross-Krook (BGK) equation \cite{Nouri} \cite{Yano2}. On the other hand, two numerical methods to solve the U-U equation were proposed \textcolor{black}{in previous studies}. One is the direct simulation Monte Carlo (DSMC) method by Garcia and Wagner \cite{Garcia} \cite{Bird}. The other is the spectral method on the basis of the Fourier transformation of the velocity distribution function by Filbet \textit{et al}. \cite{Filbet}. The DSMC method by Garcia and Wagner \cite{Garcia} requires a markedly large number of sample particles and lattices in the velocity space ($\mathbb{V}^3 \subseteq \mathbb{R}^3$) to reproduce the accurate thermalization \textcolor{black}{(equilibration)}, in other words, \textcolor{black}{directivity} toward the thermally equilibrium distribution, namely, Bose-Einstein (B.E.) distribution or Fermi-Dirac (F.D.) distribution \textcolor{black}{as a result of binary collisions}. Similarly, the spectral method by Filbet \textit{et al}. \cite{Filbet} requires lattices in $\mathbb{K}^3$ ($\mathbb{K}^3$ is the wave number space as a result of the Fourier transformation of $\mathbb{V}^3$). Finally, the spectral method also requires six dimensional lattices to express the distribution function $f\left(\bm{k},\bm{x}\right)$ in $\mathbb{K}^3 \times \mathbb{X}^3$ ($\bm{x} \in \mathbb{X}^3$: physical space, $\bm{k} \in \mathbb{K}^3$) as well as the DSMC method by Garcia and Wagner \cite{Garcia}, when we calculate three dimensional flow. Additionally, the spectral method by Filbet \textit{et al}. \cite{Filbet} does not always satisfy the positivity of the velocity distribution function, namely, $\exists f\left(\bm{k},\bm{x}\right) \notin \mathbb{R}^+$ owing to the characteristics of the Fourier transformation. Consequently, the accurate calculation of the dilute quantum gas on the basis of \textcolor{black}{previous} two numerical methods has been difficult for us even with the most advanced high performance parallel computers. Similarly, the computation of the quantum kinetic equation such as the quantum FPE or quantum BGK equation also requires parallel computers to solve the $d+3$ dimensional distribution function ($\mathbb{V}^3 \times \mathbb{X}^d$, $d=2,3$), when we calculate $d$-dimensional flow of the dilute quantum gas. \textcolor{black}{Additionally, the quantum FPE includes the numerical instability owing to its mathematical structure of the collisional term \cite{Yano1}, whereas the quantum BGK model cannot express the nonlinearity involved with the collisional term in the U-U model equation and postulates the undermined relaxational time, which must be determined from the viscosity coefficient derived from the U-U equation.} In this paper, the DSMC method is considered to calculate the dilute quantum gas, accurately, in practical time by modifying the collisional term in the U-U equation. As described by Garcia and Wagner \cite{Garcia}, the primary difficulty involved with the calculation of the U-U equation on the basis of the DSMC method is the calculation of the velocity distribution function after the binary collision, which requires fine lattices in $\mathbb{V}^3$ to reproduce the B.E. or F.D. distribution as a steady solution of the velocity distribution function under the spatially homogeneous state. From numerical results by Garcia and Wagner \cite{Garcia}, $N=10^6$ sample particles and $100^3$ lattices in $\mathbb{V}^3$ are required to reproduce the B. E. distribution, accurately. Thus, we propose the U-U model equation by assuming that two distribution functions in the collisional term of the U-U equation, which determine the collision frequency, can be modified with the thermally equilibrium distribution function, namely, F.D. or B.E. distribution. As a result of such a modification of the collisional term in the U-U equation, we need not to use fine lattices in $\mathbb{V}^3$ to calculate the distribution function after the binary collision. In this paper, we try to prove that $N$-dimensional flow ($N \le 2$) of the dilute and non-condensed quantum gas can be calculated in practical time, \textcolor{black}{accurately, using such a U-U model equation}.\\
In this paper, we restrict ourselves to the boson to simplify our discussion, whereas the calculation of the fermion can be readily performed using the same DSMC algorithm. \textcolor{black}{The DSMC results surely confirm that the accurate thermalization is obtained by solving the U-U model equation. The relation between the U-U model equation and U-U equation is considered by comparing the viscosity coefficients of the hard spherical molecule and (pseudo) Maxwellian molecule \cite{Cercignani}, which are obtained using the U-U model equation, with the analytical result of the viscosity coefficient, which was derived from the U-U equation by Nikuni-Griffin \cite{Nikuni}. Here, readers must remind that the viscosity coefficient by Nikuni-Griffin was obtained using the collisional cross section, which is independent of both deflection angle and relative velocity between two colliding molecules, because the collisional cross section is determined by the quantum mechanics. On the other hand, the collisional cross section of the hard spherical molecule depends on both deflection angle and relative velocity and that of the Maxwellian molecule depends on not relative velocity but deflection angle, because collisional cross sections of the hard spherical molecule and Maxwellian molecule are determined by the classical mechanics. In this paper,  collisional cross sections of the hard spherical molecule and Maxwellian molecule are investigated to approximate the viscosity coefficient of the hard spherical molecule and Maxwellian molecule to their classical values, when the fugacity approximates to zero. The U-U equation, which was used by Nikuni-Griffin \cite{Nikuni}, however, never approximates to the classical Boltzmann equation owing to its definition of the collisional cross section. Thus, comparisons of viscosity coefficients are performed for understanding of the characteristics of the viscosity coefficient, which is obtained using the U-U model equation, as a function of the fugacity and inverse power law number rather than the proof of the relevance of the U-U model equation as a kinetic model of the U-U equation.\\
To emphasize the availability of the proposed numerical method, the shock layer of the dilute Bose gas around the cylinder is investigated. The DSMC results of the shock layer certainly indicate that the proposed numerical method enables us to calculate two dimensional flow of the dilute quantum gas in practical time, accurately, whereas the stiffness of the calculation, which is caused by the approximation of the fugacity to unity in the Bose gas, must be improved in our study.}
\section{U-U model equation and its numerical method}
\textcolor{black}{In this section, the U-U model equation is proposed and the DSMC algorithm is considered to solve the U-U model equation.}
\subsection{U-U model equation}
First of all, the U-U equation is written as
\begin{eqnarray}
&&\partial_t f\left(\bm{v}\right)+\bm{v} \cdot \bm{\nabla}f \left(\bm{v}\right) \nonumber \\
&=&\int_{I_\chi \times I_\epsilon} \int_{\mathbb{V}_1^3} \left[f\left(\bm{v}^\prime\right)f\left(\bm{v}_1^\prime\right)\left(1-\theta f\left(\bm{v}\right)\right)\left(1-\theta f\left(\bm{v}\right)\right) \right. \nonumber \\
&&\left. -f\left(\bm{v}\right)f\left(\bm{v}_1\right)\left(1-\theta f\left(\bm{v}^\prime\right)\right)\left(1-\theta f\left(\bm{v}^\prime\right)\right)\right] \nonumber \\
&& g \sigma \sin \chi d \epsilon d\chi d \bm{v}_1, \nonumber \\
&& \theta=-1~ (\text{Boson}),~~~\theta=+1~ (\text{Fermion}),
\end{eqnarray}
where $f\left(\bm{v}\right):=f\left(t,\bm{v},\bm{x}\right)$ in $\mathbb{R}^+ \times \mathbb{V}^3 \times \mathbb{X}^3$ ($t$: time) is the velocity distribution function, $\bm{v}_1 \in \mathbb{V}_1^3$ is the velocity of the collisional partner, $\sigma$ is the differential cross section, $\chi \in I_{\chi}, I_\chi:=\left[0,\pi\right]$ is the deflection angle, and $\epsilon \in I_{\epsilon}$, $I_\epsilon:=\left[0,2\pi\right]$ is the scattering angle.\\
The difficulty involved with solving the U-U equation in Eq. (1) on the basis of the DSMC method by Garcia and Wagner \cite{Garcia} is caused by the evaluation of $f\left(\bm{v}^\prime\right)$ and $f\left(\bm{v}_1^\prime\right)$. In each time step, $f\left(\bm{\ell}\right)$ is calculated by counting up sample particles inside a lattice $\bm{\ell}$ in $\bm{\ell} \in \bm{V}^3$, so that $f\left(\bm{v}^\prime\right)$ is obtained using $f\left(\bm{\ell}\right)$, in which $\bm{v}^\prime \in \bm{\ell}$.\\
Garcia and Wagner \cite{Garcia} indicated that the number of sample particles must be equivalent to the number of lattices, namely, $\left|\bm{\ell}\right|$ to obtain the B.E. or F.D. distribution as a steady solution of $f\left(\bm{v}\right)$ under the spatially homogeneous state, namely, $\bm{\nabla} f\left(\bm{v}\right)=\bm{0}$ in Eq. (1).\\
To avoid the evaluation of $f\left(\bm{\ell}\right)$, we consider the U-U model equation such as
\begin{eqnarray}
&&\partial_t f\left(\bm{v}\right)+\bm{v} \cdot \bm{\nabla}f \left(\bm{v}\right) \nonumber \\
&=&\int_{I_\chi \times I_\epsilon} \int_{\mathbb{V}_1^3} \left[f\left(\bm{v}^\prime\right)f\left(\bm{v}_1^\prime\right)\left(1-\theta f^{\text{eq}}\left(\bm{v}\right)\right)\left(1-\theta f^{\text{eq}}\left(\bm{v}_1\right)\right) \right. \nonumber \\
&&\left. -f\left(\bm{v}\right)f\left(\bm{v}_1\right)\left(1-\theta f^{\text{eq}}\left(\bm{v}^\prime\right)\right)\left(1-\theta f^{\text{eq}}\left(\bm{v}^\prime_1\right)\right)\right] \nonumber \\
&& g \sigma \sin \chi d \epsilon d\chi d \bm{v}_1,
\end{eqnarray}
where the equilibrium distribution function $f^{\text{eq}}\left(\bm{v}\right)$ is defined as $f^{\text{eq}}\left(\bm{v}\right):=F\left(\bm{u},T,\mathfrak{Z}\right)=\left\{\mathfrak{Z}^{-1}\exp\left( \tilde{C}^2\right)+\theta\right\}^{-1}$, in which $\mathfrak{Z}:=\exp\left[\left(\mu\left(t,\bm{x}\right)-U\left(t,\bm{x}\right)\right)/\left(RT\right)\right]$ ($\mu$: chemical potential, $U$: effective potential) is the fugacity and $\tilde{\bm{C}}:=\bm{C}/\sqrt{2RT}$ ($R=k/m$: gas constant, $k$: Boltzmann constant, $m$: mass of a molecule, $T$: temperature: $\bm{C}:=\bm{v}-\bm{u}$: peculiar velocity \cite{Cercignani}, $\bm{u}$: flow velocity). Macroscopic quantities, which define $f^{\text{eq}}\left(\bm{v}\right)$, are calculated by $\rho=\left(\hat{h}^3/m\right)\int_{\mathbb{V}^3} f\left(\bm{v}\right) d\bm{v}$, $\rho \bm{u}=\left(\hat{h}^3/m\right)\int_{\mathbb{V}^3}\bm{v} f\left(\bm{v}\right) d\bm{v}$ and $p=(1/3)\left(\hat{h}^3/m\right)\int_{\mathbb{V}^3}C^2 f\left(\bm{v}\right) d\bm{v}$, where $\rho$ is the density and $\hat{h}=h/m$ ($h$: Planck constant).\\
\textcolor{black}{The U-U model equation in Eq. (2) is straightforwardly derived from the U-U equation by expanding $f\left(\bm{v}\right)$ ($f\left(\bm{v}_1\right)$) around $f^{\text{eq}}\left(\bm{v}\right)$ ($f^{\text{eq}}\left(\bm{v}_1\right)$) such as $f\left(\bm{v}\right)=f^{\text{eq}}\left(\bm{v}\right)\left(1+\phi\left(\bm{v}\right)\right)$ ($f\left(\bm{v}_1\right)=f^{\text{eq}}\left(\bm{v}_1\right)\left(1+\phi\left(\bm{v}_1\right)\right)$) and $f\left(\bm{v}^\prime\right)$ ($f\left(\bm{v}_1^\prime\right)$) around $f^{\text{eq}}\left(\bm{v}^\prime\right)$ ($f^{\text{eq}}\left(\bm{v}_1^\prime\right)$) such as $f\left(\bm{v}^\prime\right)=f^{\text{eq}}\left(\bm{v}^\prime\right)\left(1+\phi\left(\bm{v}^\prime\right)\right)$ ($f\left(\bm{v}_1^\prime\right)=f^{\text{eq}}\left(\bm{v}_1^\prime\right)\left(1+\phi\left(\bm{v}_1^\prime\right)\right)$) in terms $\left[1-\theta f\left(\bm{v}\right)\right]\left[1-\theta f\left(\bm{v}_1\right)\right]$ and $\left[1-\theta f\left(\bm{v}^\prime\right)\right]\left[1-\theta f\left(\bm{v}_1^\prime\right)\right]$ in the right hand side of Eq. (1), in which $f^{\text{eq}} \phi $ is the deviation from the thermally equilibrium distribution, namely, F. D. or B. E. distribution, and taking the zeroth order approximation, namely, $\phi=0$. Of course, the U-U model equation in Eq. (2) can be improved by approximating $\phi$ with Grad's 13 moments, as discussed in appendix B.}
\subsection{Numerical method to solve U-U model equation}
The DSMC algorithm is considered to calculate the U-U model equation in Eq. (2). As a collisional scheme, the majorant collision frequency by Ivanov \cite{Ivanov} is applied. The majorant collision frequency $\nu_{\text{max}}$ is calculated for the variable hard sphere (VHS) \cite{Bird}, which demonstrates the molecule with inverse power law potential (IPL), such as
\begin{eqnarray}
\nu_{\max}&=&\left[\left(1-\theta f^{\text{eq}}\left(\bm{v}^\prime\right)\right)\left(1-\theta f^{\text{eq}}\left(\bm{v}_1^\prime\right)\right)\right]_{\max} g_{\max}^\xi \mathfrak{A} \nonumber \\
&=& \left(\frac{\mathfrak{Z}^{-1}}{\mathfrak{Z}^{-1}-1}\right)^2 g_{\max}^\xi \mathfrak{A}~~~(\theta=-1:~\text{boson}), \nonumber \\
&=& g_{\max}^\xi \mathfrak{A}~~~(\theta=+1:~\text{fermion}),
\end{eqnarray}
where $\mathfrak{A}=1/2 \sigma_T n \left(N-1\right) \Delta t$ ($n$: number density, $N$: number of sample particles in a lattice, $\sigma_T$: total cross section, $\Delta t$: time interval) and $\xi\in \left[0,1\right]$ ($\xi=0$: pseudo Maxwellian molecule, $\xi=1$: hard sphere molecule).\\
Once $\nu_{\max}$ is calculated, collisional pairs are chosen $\nu_{\max}$ times. The collision occurs, when
\begin{eqnarray}
\mathscr{W}_1<\frac{g^\xi}{g_{\max}^\xi} \wedge \mathscr{W}_2<\frac{\left(1-\theta f^{\text{eq}}\left(\bm{v}^\prime \right)\right)\left(1-\theta f^{\text{eq}}\left(\bm{v}_1^\prime\right)\right)}{\left[\left(1-\theta f^{\text{eq}}\left(\bm{v}^\prime\right)\right)\left(1-\theta f^{\text{eq}}\left(\bm{v}_1^\prime\right)\right)\right]_{\max}}
\end{eqnarray}
where $\mathscr{W}_1, ~\mathscr{W}_2 \in \left[0,1\right]$ is the white noise.\\
\textcolor{black}{From Eq. (3), Knudsen number (Kn) for the boson is defined as}
\begin{eqnarray}
\text{Kn}=\text{Kn}|_{\mathfrak{Z}\rightarrow 0} \left(1-\mathfrak{Z}\right)^2,
\end{eqnarray}
\textcolor{black}{where $\text{Kn}|_{\mathfrak{Z}\rightarrow 0}$ is Knudsen number, which is calculated using the mean free path for the classical gas ($\mathfrak{Z}\rightarrow 0$).}\\
Above numerical scheme is markedly simpler than the DSMC method by Garcia and Wagner \cite{Garcia}. The only difficulty in the calculation of the U-U model equation is that $\nu_{\max}$ for the boson increases, as $\mathfrak{Z}$ approximates to unity. \textcolor{black}{In other words, Kn approximates zero, as $\mathfrak{Z}$ approximates to unity, from Eq. (5).} As a result, the computational time increases, as $\mathfrak{Z}$ of the boson approximates to unity.\\
\textcolor{black}{In this paper, all the physical quantities are normalized. The quantities with $\sim$ are normalized such as $\tilde{\rho}:=\rho/\rho_\infty$, $\tilde{\bm{u}}:=\bm{u}/C_\infty$, $\tilde{\bm{v}}:=\bm{v}/C_\infty$, ($C_\infty:=\sqrt{2RT_\infty}$),  $\tilde{T}:=T/T_\infty$, $\tilde{\bm{x}}:=\bm{x}/L_\infty$, $\tilde{t}:=t/\left(L_\infty/C_\infty\right)$ and $\tilde{f}:=f/\left(\pi^{3/2}\text{Li}_{2/3}(\mathfrak{Z}_\infty)\right)$.\\
In the DSMC method, $\tilde{\rho}_{i,j,k}$, $\tilde{\bm{u}}_{i,j,k}$, $\mathfrak{Z}_{i,j,k}$ and $\tilde{T}_{i,j,k}$ in the lattice $\left(i,j,k\right)$ is calculated for the boson such as}
\begin{eqnarray}
&&\tilde{\rho}_{i,j,k}:=N_{i,j,k}/\left(N_s \left|V_{i,j,k}\right|\right),~~~N_{i,j,k}:=\left|\bigcup_{\bm{x}_\ell \in V_{i,j,k}} \ell \right|\\
&&\tilde{\bm{u}}_{i,j,k}:=\sum_{\ell |\bm{x}_\ell \in V_{i,j,k}} \tilde{\bm{v}}_\ell /N_{i,j,k},\\
&&\frac{2}{3}\sum_{\ell |\bm{x}_\ell \in V_{i,j,k}} \left(\tilde{\bm{v}}_\ell-\tilde{\bm{u}}_{i,j,k}\right)^2 /N_{i,j,k}=\frac{\tilde{\rho}_{i,j,k}^{\frac{2}{3}}}{\left[\text{Li}_{\frac{3}{2}}\left(\mathfrak{Z}_{i,j,k}\right)\right]^{\frac{5}{3}}}\frac{\text{Li}_{\frac{5}{2}}\left(\mathfrak{Z}_{i,j,k}\right)}{\left[\text{Li}_{\frac{3}{2}}\left(\mathfrak{Z}_\infty\right)\right]^{\frac{2}{3}}},\\
&&\tilde{T}_{i,j,k}:=\left[\tilde{\rho}_{i,j,k} \frac{\text{Li}_{\frac{2}{3}}\left(\mathfrak{Z}_\infty\right)}{\text{Li}_{\frac{2}{3}}\left(\mathfrak{Z}_{i,j,k}\right)}\right]^{\frac{2}{3}},
\end{eqnarray}
\textcolor{black}{where $V_{i,j,k} \in \mathbb{X}^3$ is the physical domain occupied by the lattice with the address $\left(i,j,k\right)$, $|V_{i,j,k}|$ is the volume of the lattice with the address $\left(i,j,k\right)$, $N_{i,j,k}$ is the number of sample particles, which are included in the lattice with the address $\left(i,j,k\right)$, $N_s$ is the number of sample particles, which corresponds to the number density, namely, $\rho_\infty/m$, $\ell \in \mathbb{N}$ is the $\ell$-th sample particle in the calculation domain, $\text{Li}\left(-\mathfrak{Z}\theta\right)$ is polylogarithm. Finally, the order of the calculation is (6) $\rightarrow$ (7) (calculation of $\tilde{\bm{u}}_{i,j,k}$ from $\tilde{\rho}_{i,j,k}$ in Eq. (6)) $\rightarrow$ (8) (calculation of $\mathfrak{Z}_{i,j,k}$ from $\tilde{\rho}_{i,j,k}$ in Eq. (6) and $\tilde{\bm{u}}_{i,j,k}$ in Eq. (7)) $\rightarrow$ (9) (calculation of $\tilde{T}_{i,j,k}$ from $\tilde{\rho}_{i,j,k}$ in Eq. (6) and $\mathfrak{Z}_{i,j,k}$ in Eq. (8))}.
\section{Numerical results}
Hereafter, we examine the accuracy of our DSMC algorithm and confirm the practical usefulness of the U-U model equation, which is solved by the DSMC method.\\
\subsection{Numerical confirmation of H theorem}
The significant condition, which must be satisfied by the simplified U-U equation, is H theorem. \textcolor{black}{H theorem in the U-U model equation is discussed in appendix A, theoretically}. Here, we try to confirm that H theorem is satisfied, when the U-U model equation is calculated using the proposed DSMC method. The U-U model equation has the practical advantage over the U-U equation, when the accurate thermalization is obtained by solving the U-U model equation using a small number of sample particles.\\
As initial datum, two types of $f\left(0,\bm{v}\right)$, namely Tests A and B, are considered. $f\left(0,\bm{v}\right)$ is set as $f\left(0,\bm{v}\right)=F\left(\bm{u}_A,T_A,\mathfrak{Z}_A\right)$ in Test A, and $f\left(0,\bm{v}\right)=\left[F\left(\bm{u}_B,T_B,\mathfrak{Z}_B\right)+F\left(-\bm{u}_B,T_B,\mathfrak{Z}_B\right)\right]/2$ in Test B, where $\mathfrak{Z}_A=0.95$, $\tilde{\bm{u}}_A=\left(1,0,0\right)$ and $\tilde{T}_A=1$ in Test A and $\mathfrak{Z}_B=0.9$, $\tilde{\bm{u}}_B=\left(1,0,0\right)$ and $\tilde{T}_B=1$ in Test B. Additionally, $\xi=0$, namely, (pseudo) Maxwellian molecule is considered, and number of sample particles are set as $N=100$ in unit lattice, which is a small number of sample particles in the DSMC calculation.\\
$f\left(0,\bm{v}\right)$ must be invariant during the time evolution in Test A, whereas $f\left(0,\bm{v}\right)$ must change toward $f^{\text{eq}}\left(\bm{v}\right)$ in Test B. Of course, quantum effects via the spin ($\theta=\pm 1$) becomes significant, as $\mathfrak{Z}$ approximates to unity, whereas quantum effects via the spin ($\theta=\pm 1$) becomes weak, as $\mathfrak{Z}$ approximates to zero. Thus, the reproduction of $f^{\text{eq}}\left(\bm{v}\right)$ requires the more accurate evaluation of $\left(1-\theta f(\bm{v})\right)\left(1-\theta f\left(\bm{v}_1\right)\right)$ in the DSMC algorithm by Garcia and Wagner \cite{Garcia}, as $\mathfrak{Z}$ approximates to unity. The enhancement of the accuracy in the DSMC algorithm by Garcia and Wagner attributes to the increases of $N$ and number of lattices in $\mathbb{V}^3$, which means the marked increase of the computational source.
\begin{figure}
\includegraphics[width=1.0\linewidth]{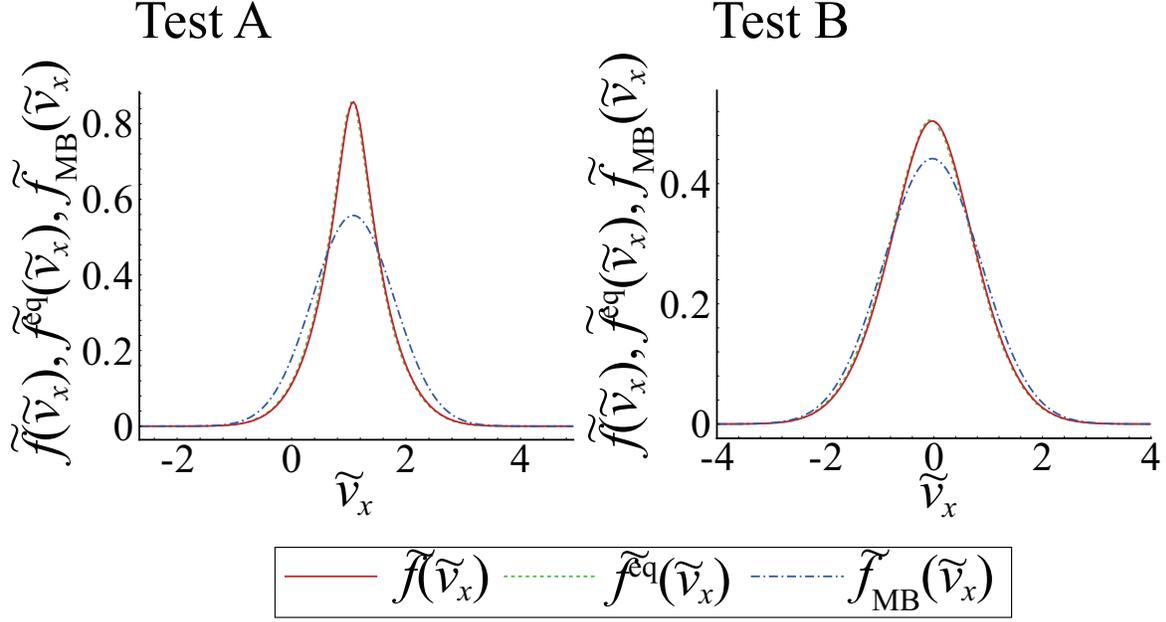}\\
\caption{$\tilde{f}\left(\tilde{v}_x\right)$ versus $\tilde{v}_x$ together with $\tilde{f}^{\text{eq}}\left(\tilde{v}_x\right)$ versus $\tilde{v}_x$ and $\tilde{f}_{\text{MB}}\left(\tilde{v}_x\right)$ versus $\tilde{v}_x$ in Tests A (left frame) and B (right frame).}
\end{figure}
Figure 1 shows $\tilde{f}\left(\tilde{v}_x\right)$ versus $\tilde{v}_x$ together with $\tilde{f}^{\text{eq}}\left(\tilde{v}_x\right)$ versus $\tilde{v}_x$ and $\tilde{f}_{\text{MB}}\left(\tilde{v}_x\right)$ versus $\tilde{v}_x$ in Tests A (left frame) and B (right frame), in which $\tilde{f}\left(\tilde{v}_x\right):=\tilde{\rho}^{-1} \int_{\mathbb{V}^2} \tilde{f}\left(\tilde{\bm{v}}\right) d\tilde{v}_y d\tilde{v}_z$, $\tilde{f}^{\text{eq}}\left(\tilde{v}_x\right):=\tilde{\rho}^{-1} \int_{\mathbb{V}^2} \tilde{f}^{\text{eq}}\left(\tilde{\bm{v}}\right) d\tilde{v}_y d\tilde{v}_z$ and $\tilde{f}_{MB}\left(\bm{v}\right):=\tilde{\rho}^{-1} \int_{\mathbb{V}^2} \tilde{f}_{MB}\left(\bm{v}\right)d\tilde{v}_y d\tilde{v}_z$ ($ \tilde{f}_{MB}\left(\bm{v}\right)$: Maxwell-Boltzmann distribution). $\tilde{f}\left(\tilde{\bm{v}}\right)$ is calculated by the ensemble average at $0 \le \tilde{t}$ in Test A and at $\tilde{t}_{\text{st}}\le \tilde{t}$ in Test B, in which $\tilde{t}_{\text{st}}$ corresponds to the time of the steady state. The number of iterations is $2\times 10^4$ in total, $\text{Kn}|_{\mathfrak{Z}\rightarrow 0}=0.01$, $\tilde{V}=1$ ($V$: volume of lattice) and $\Delta \tilde{t}=2.5\times10^{-2}$, so that the total CPU time is about forty minutes in Test A, or five minutes in Test B, using 2.5 GHz processor. As shown in the left and right frames, $\tilde{f}\left(\tilde{v}_x\right) \simeq \tilde{f}^{\text{eq}}\left(\tilde{v}_x\right)$ is obtained in both Tests A and B, so that H theorem in the U-U model equation is proved, numerically. Of course, $\tilde{f}\left(\tilde{v}_x\right)$ is clearly different from $\tilde{f}_{\text{MB}}\left(\tilde{v}_x\right)$ \textcolor{black}{in both Tests A and B}. Above numerical results also show that our DSMC algorithm enables us to calculate the U-U model equation in practical time, whereas the increase of $\mathfrak{Z}$ surely yields the increase of the total CPU time owing to the increase of the majorant collision frequency of bosons in Eq. (3).
\subsection{Numerical result of viscosity coefficient by U-U model equation}
Next, we consider the viscosity coefficient $(\eta)$ of bosons, which is obtained using the U-U model equation. The kinetic calculation of the viscosity coefficients of bosons, which is obtained using the U-U model equation, is as difficult as that obtained using the U-U equation. Then, we numerically investigate the viscosity coefficient of bosons, whose intermolecular potential is described by the IPL potential. We obtain the time evolution of the time-correlation function of the pressure deviator on the basis of the two-point kinetic theory by Tsuge and Sagara \cite{Tsuge} and Grad's 13 moment equation \cite{Grad} for the quantum gas, which was calculated by the author \cite{Yano1}, such as
\begin{eqnarray}
\frac{d Q_{ij,kl}^{(2,2)}(\tau)}{d\tau}=-\frac{p}{\eta}Q_{ij,kl}^{(2,2)}\left(\tau\right).
\end{eqnarray}
where
\begin{eqnarray} 
Q_{ij,kl}^{(2,2)}(\tau)&:=&\left<\int_{\mathbb{V}^3} H_{ij}^{(2)}\left(t_1,\bm{v}\right) f\left(t_1,\bm{v}\right) d\bm{v}, \right. \nonumber \\
&&\left. \int_{\mathbb{V}^3} H_{kl}^{(2)}\left(t_2,\bm{v}\right) f\left(t_2,\bm{v}\right) d\bm{v}\right>,
\end{eqnarray}
in which $H_{ij}^{(2)}=\tilde{C}_i\tilde{C}_j-\delta_{ij} \tilde{C}^2/3$ and $\tau=t_2-t_1$. Of course, $<,>$ reveals the ensemble average.
From Eq. (10), we obtain
\begin{eqnarray}
\eta=p\int_0^\infty Q_{ij,kl}^{(2,2)}(\tau) d\tau \left[Q_{ij,kl}^{(2,2)}(0)\right]^{-1}.
\end{eqnarray}
Gust and Reichl \cite{Gust} calculated transport coefficients for the hard sphere (HS) molecule ($\xi=1$ in Eq. (3)) using Rayleigh-Schr$\ddot{\text{o}}$dinger perturbation theory rather than the kinetic calculation on the basis of Chapman-Enskog method \cite{Chapman}. The viscosity coefficient, which was calculated by Gust and Reichl, does not approximate to the first order Chapman-Enskog approximation of the viscosity coefficient of the classical gas, when $\mathfrak{Z} \rightarrow 0$. Meanwhile, Nikuni and Griffin calculated transport coefficients of the Bose gas by numerically calculating the collisional term of the U-U equation, whereas the U-U equation, which was considered by Nikuni and Griffin, postulates the constant collisional cross section, which is independent of the relative velocity of two colliding molecules and deflection angle, namely, $g\sin \chi$, in the right hand side of Eq. (1). Therefore, the form of the U-U equation, which was studied by Nikuni and Griffin, is clearly different from Eq. (1). We, however, compare the viscosity coefficient derived from the U-U model equation, which is calculated by Eq. (12) using the DSMC method, with that obtained by Nikuni and Griffin together with the viscosity coefficient derived from the quantum BGK equation, because nobody has succeeded the kinetic calculation of the transport coefficients derived from the U-U equation in Eq. (1). The quantum BGK equation is written as $\partial_t f+ \bm{v} \cdot \bm{\nabla}f=\left(f^{\text{eq}}-f\right)/\mathfrak{T}$, where $\mathfrak{T}$ is the relaxation time. As the initial numerical condition to calculate Eq. (2), $5000$ sample particles are set in a unit lattice, where equally spaced $5 \times 5$ lattices are set in the square domain $x\in[0,1]$ and $y \in [0,1]$, $\tilde{\rho}|_{\tilde{t}=0}=\tilde{T}|_{\tilde{t}=0}=1$, and $\tilde{\bm{u}}|_{\tilde{t}=0}=(0,0,0)$. The initial fugacity is set as $\mathfrak{Z}|_{\tilde{t}=0}=0.01$ and $0.1\times i$ $(1\le i \le 9 \cap i \in \mathbb{N})$. For convenience, we use the numerical result of $\mathfrak{Z}|_{\tilde{t}=0}=0.01$ as the value under the classical limit, namely, $\mathfrak{Z}\rightarrow 0$, because thermal characteristics at $\mathfrak{Z}|_{\tilde{t}=0}=0.01$ are presumably similar to those at $\mathfrak{Z}|_{\tilde{t}=0}=0$. The time interval is set as $\Delta \tilde{t}=2.5 \times 10^{-5}$ and total iteration number is $2.0 \times 10^5$. We calculate two types of the VHS molecule, namely, the HS molecule and (pseudo) Maxwellian molecule. $\text{Kn}|_{\mathfrak{Z}\rightarrow 0}=2.5 \times 10^{-4}$ for the Maxwellian molecule and $\text{Kn}|_{\mathfrak{Z}\rightarrow 0}=2.5(2)^{-1/2} \times 10^{-4}$ for the HS molecule, respectively. The heaviest calculation requires about 24 hours using 3.0 GHz processor in the case of the HS molecule with $\mathfrak{Z}|_{\tilde{t}=0}=0.9$.
\begin{figure}
\includegraphics[width=0.9\linewidth]{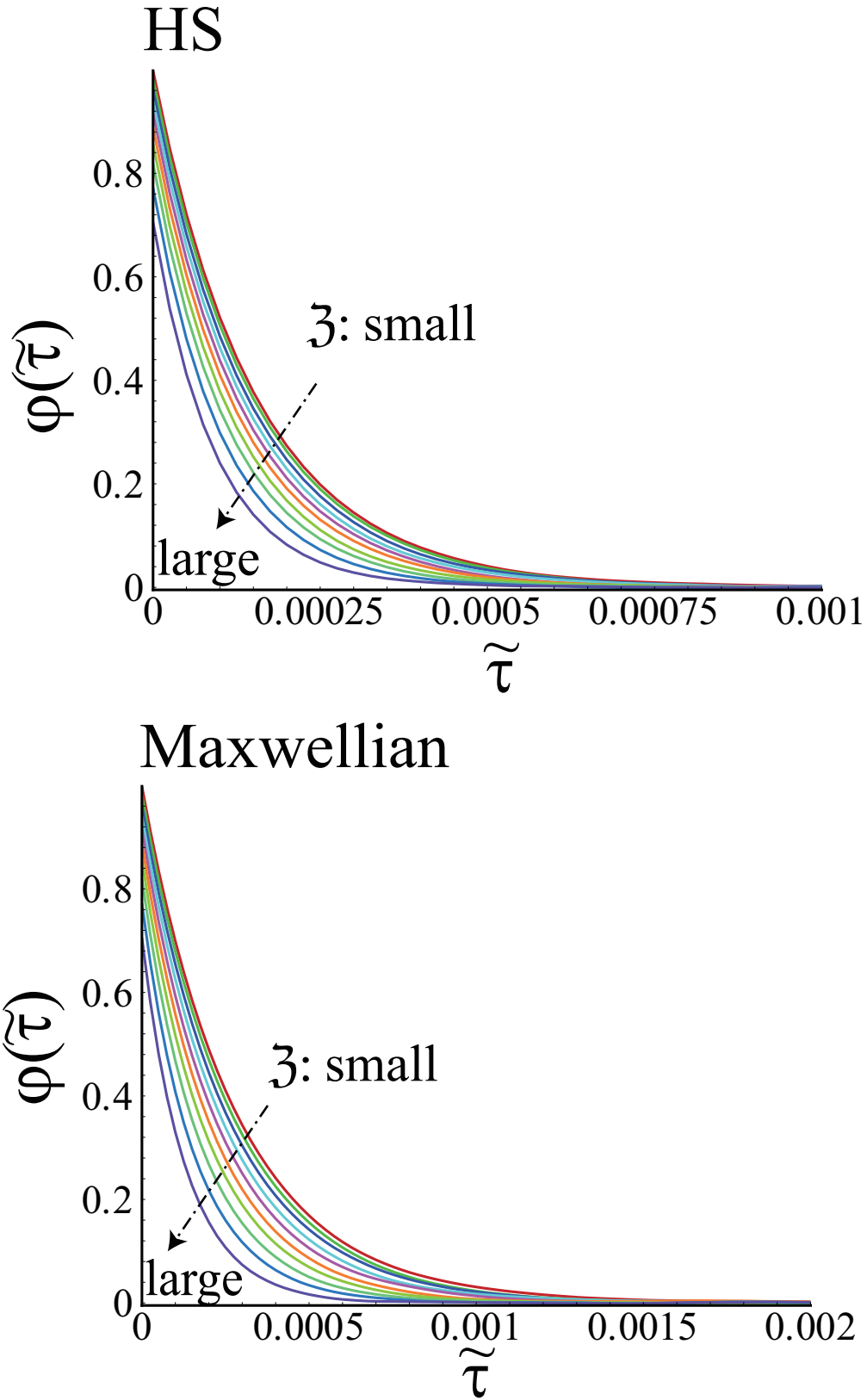}\\
\caption{$\varphi\left(\tilde{\tau}\right)$ versus $\tilde{\tau}$ for the HS (upper frame) and Maxwellian molecule (lower frame), when $\mathfrak{Z}|_{\tilde{t}=0}=0.01$. $0.1\times i$ ($1\le i \le 9 \cap i \in \mathbb{N}$)}
\end{figure}
Figure 2 shows $\varphi(\tilde{\tau}):=\tilde{p} Q_{xx,xx}^{(2,2)}\left(\tilde{\tau}\right)/Q_{xx,xx}^{(0,0)}\left(0\right)$ versus $\tilde{\tau}$ for the HS molecule (upper frame) and Maxwellian molecule (lower frame). From $\varphi\left(0\right)=\tilde{p}$, $\varphi\left(0\right)$ decreases, as $\mathfrak{Z}|_{\tilde{t}=0}$ decreases, as confirmed in the upper and lower frames of Fig. 2, because $p$ decreases, as $\mathfrak{Z}$ decreases. The upper and lower frames of Fig. 2 show that the damping rate of $\varphi(\tilde{\tau})$ increases, as $\mathfrak{Z}|_{\tilde{t}=0}$ decreases. Of course, the damping rate of $\varphi(\tilde{\tau})$ in the case of the HS molecule is larger than that in the case of the Maxwellian molecule, because Kn for the HS molecule is smaller than that for the Maxwellian molecule. The normalized viscosity coefficient is obtained by integrating $\varphi(\tilde{\tau})$ in $\tilde{\tau} \in [0,\infty]$, whereas such an integration corresponds to the area surrounded by $x$ and $y$ axises and $\varphi\left(\tilde{\tau}\right)$. In this paper, $\eta\left(\mathfrak{Z}\right)/\eta\left(0\right)$ is calculated to compare our DSMC results with the previous result of $\eta\left(\mathfrak{Z}\right)/\eta\left(0\right)$ by Nikuni and Griffin \cite{Nikuni}. Additionally, $\eta\left(\mathfrak{Z}\right)/\eta\left(0\right)$, which is calculated by the quantum BGK equation, is considered. From the author's previous study \cite{Yano1}, $\eta\left(\mathfrak{Z}\right)/\eta\left(0\right)$ for the quantum BGK equation is calculated as $\eta\left(\mathfrak{Z}\right)/\eta\left(0\right)=p\left(\mathfrak{Z}\right)/p\left(0\right)$, when $\mathfrak{T}$ is assumed to be independent of $\mathfrak{Z}$.
\begin{figure}
\includegraphics[width=1.0\linewidth]{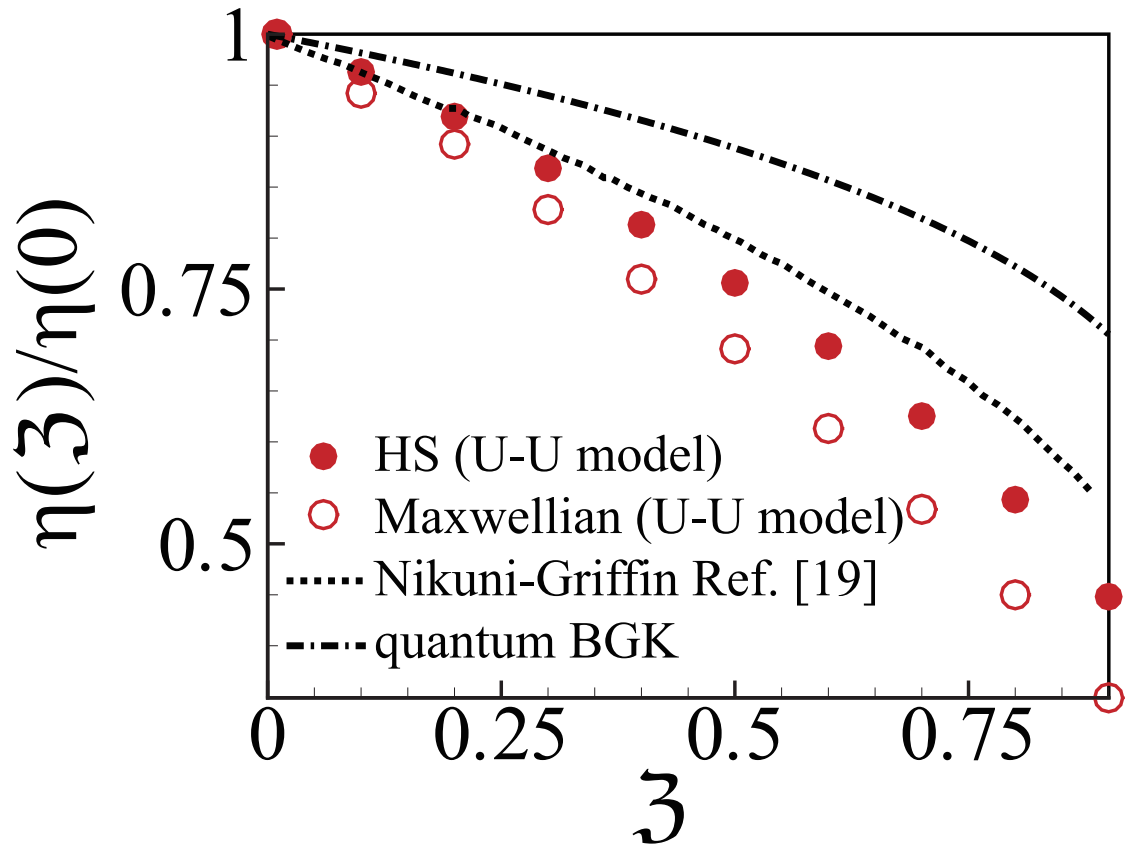}\\
\caption{$\eta\left(\mathfrak{Z}\right)/\eta(0)$ versus $\mathfrak{Z}$ for the HS and Maxwellian molecule, which are calculated using the U-U model equation together with $\eta\left(\mathfrak{Z}\right)/\eta(0)$ versus $\mathfrak{Z}$, which was calculated by Nikuni and Griffin \cite{Nikuni}, and $\eta\left(\mathfrak{Z}\right)/\eta(0)$ versus $\mathfrak{Z}$, which is calculated by the quantum BGK equation with fugacity independent relaxation time.}
\end{figure}
Figure 3 shows $\eta\left(\mathfrak{Z}\right)/\eta\left(0\right)$ for the HS and Maxwellian molecules together with $\eta\left(\mathfrak{Z}\right)/\eta\left(0\right)$ by Nikuni and Griffin \cite{Nikuni} and $\eta\left(\mathfrak{Z}\right)/\eta\left(0\right)$ for the quantum BGK equation. Firstly, we find that the inclination of $\eta\left(\mathfrak{Z}\right)/\eta\left(0\right)$ for the Maxwellian molecule is smallest, whereas the inclination of $\eta\left(\mathfrak{Z}\right)/\eta\left(0\right)$ for the quantum BGK equation is largest. In particular, $\eta\left(\mathfrak{Z}\right)/\eta\left(0\right)$ depends on $\xi$ (form of the intermolecular potential), because $\eta\left(\mathfrak{Z}\right)/\eta\left(0\right)$ for the HS molecule is different from that for the Maxwellian molecule. Surely, we can confirm that the rate of the increase of the damping rate in accordance with the increase of $\mathfrak{Z}$ for the Maxwellian molecule is larger than that for the HS molecule, as shown in Fig. 2. We must answer to the question why $\eta\left(\mathfrak{Z}\right)/\eta\left(0\right)$ depends on $\xi$ in our future study, furthermore. $\eta\left(\mathfrak{Z}\right)/\eta\left(0\right)$ for the HS molecule is similar to $\eta\left(\mathfrak{Z}\right)/\eta\left(0\right)$ by Nikuni and Griffin in the range of $0 \le \mathfrak{Z} \le 0.2$, whereas the difference $\eta\left(\mathfrak{Z}\right)/\eta\left(0\right)$ for the HS molecule and $\eta\left(\mathfrak{Z}\right)/\eta\left(0\right)$ by Nikuni and Griffin increases, as $\mathfrak{Z}$ increases from $\mathfrak{Z}=0.2$. $\eta\left(\mathfrak{Z}\right)/\eta\left(0\right)$ for the quantum BGK equation becomes markedly different from $\eta\left(\mathfrak{Z}\right)/\eta\left(0\right)$ for the HS and Maxwellian molecules and $\eta\left(\mathfrak{Z}\right)/\eta\left(0\right)$ by Nikuni and Griffin, as $\mathfrak{Z}$ increases, whereas the correct relaxation time $\mathfrak{T}$ in the quantum BGK equation must be a function of $\mathfrak{Z}$. Finally, the calculation of $\eta\left(\mathfrak{Z}\right)/\eta\left(0\right)$ in the range of $0.9< \mathfrak{Z}<1$ in practical time requires parallel computers owing to the marked increase of the majorant collision frequency, $\nu_{\max}$, under $\mathfrak{Z} \rightarrow 1$, as shown in Eq. (3).\\
The transport coefficients are significant to demonstrate the dissipation process of the quantum gas, accurately, whereas the transport coefficients, which are obtained using the U-U model equation in Eq. (2), are presumably different from those obtained using the U-U model equation, as discussed in appendix A. The author, however, believes that the accurate thermalization via the U-U model equation is rather significant than obtaining the correct transport coefficients, because the DSMC method for the U-U equation usually violates the accurate thermalization owing to difficulties of the accurate reproduction of $f\left(\bm{v}^\prime\right)$ ($f\left(\bm{v}_1^\prime\right)$), unless so many sample particles and lattices are used, as pointed by Garcia and Wagner \cite{Garcia}. Additionally, the author is skeptic about the use of the quantum BGK model, because nobody succeeded the calculation of the relaxation rate, namely, $\mathfrak{T}\left(\mathfrak{Z}\right)=\eta/p$, whereas we face to the instability of the quantum Fokker-Planck equation, as discussed by the author \cite{Yano2}. Finally, the author considers that the U-U model equation in Eq. (2) is the best choice owing to the accurate thermalization, when we calculate the dilute quantum gas in practical time. In terms of the reproduction of the accurate transport coefficients, further modification of the U-U model is expected on the basis of Grad's 13 moment expansion of the distribution function, as discussed in appendix B.
\subsection{Shock layer of dilute Bose gas}
\textcolor{black}{Finally, we investigate whether our numerical method enables us to calculate the dilute two dimensional flow in practical time, accurately, even when the flow-field includes the strongly nonequilibrium regime. Then, the shock layer of the dilute Bose gas is calculated using the U-U model equation on the basis of the DSMC method. The shock layer is formed around the circular cylinder, whose radius is set as $\tilde{R}=1$, as shown in Fig. 4. The center axis of the cylinder coincides with $Z$-axis. The outer boundary of the calculation domain is set as $\sqrt{\tilde{X}^2+\tilde{Y}^2}=6$ and the inner boundary is set as $\sqrt{\tilde{X}^2+\tilde{Y}^2}=1$, whereas, the forward the cylinder, namely, $\tilde{X} \le 0$ is calculated, as shown in Fig. 4.}
\begin{figure}
\includegraphics[width=0.75\linewidth]{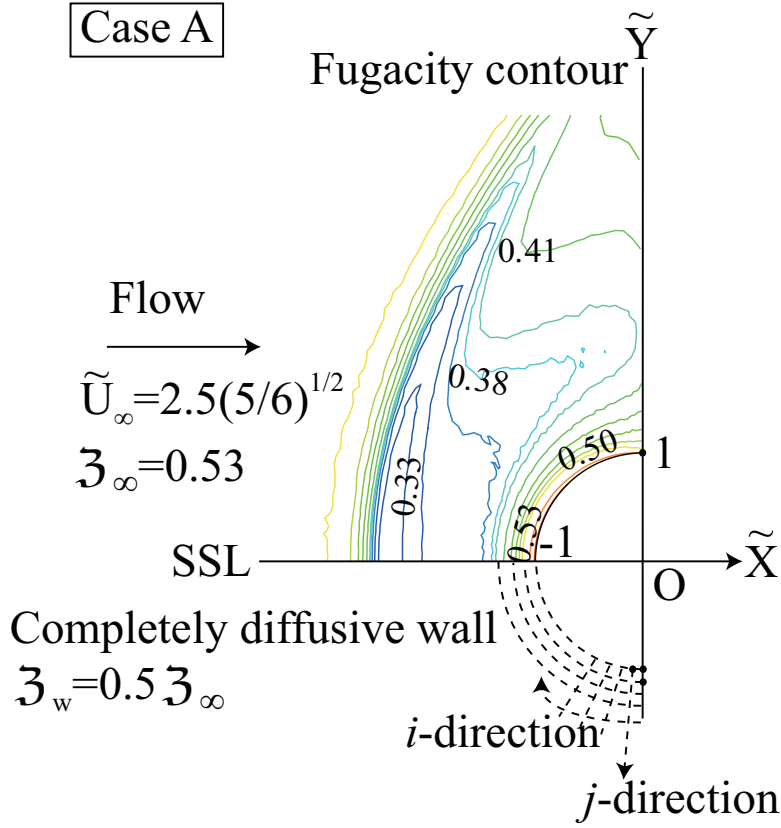}\\
\caption{Schematic of the flow field and contour of fugacity in Case A.}
\end{figure}
\textcolor{black}{Two types of the uniform flow condition, namely, Cases A and B, are considered such as}
\begin{eqnarray*}
&&\tilde{\rho}_\infty=1,~~\tilde{T}_\infty=1,~~\tilde{\bm{u}}_\infty=\left(\tilde{U}_\infty,0,0\right)=\left(2.5\sqrt{5/6},0,0\right),~~\mathfrak{Z}_w=\mathfrak{Z}_\infty/2,\\
&&\underline{\text{Case A}}\\
&&\mathfrak{Z}_\infty=0.53, ~~ M_\infty=2.34,~~\left(C_V\right)_\infty=1.59R,\\
&&\underline{\text{Case B}}\\
&&\mathfrak{Z}_\infty=0.11,~~ M_\infty=2.47,~~\left(C_V\right)_\infty=1.51R,
\end{eqnarray*}
\textcolor{black}{where quantities with $\infty$ corresponds to physical quantities in the uniform flow. Additionally, we consider the completely diffusive wall \cite{Cercignani}, whose temperature and fugacity are expressed with $T_w$ and $\mathfrak{Z}_w$, respectively. $C_V:=(15/4)\text{Li}_{5/2}\left(\mathfrak{Z}\right)/\text{Li}_{3/2}\left(\mathfrak{Z}\right)-(9/4)\text{Li}_{5/2}\left(\mathfrak{Z}\right)/\text{Li}_{3/2}\left(\mathfrak{Z}\right)$ is the specific heat at the constant volume. $\Delta \tilde{t}=0.1 \sqrt{\pi}  \text{Kn}_\infty|_{\mathfrak{Z} \rightarrow 0}$ is used in Cases A and B. Total iteration number is 6000 and ensemble averages are calculated, when the iteration number is larger than 3000. The CPU time for $\mathfrak{Z}_\infty=0.8$ is about 24 hours in Case A and 14 hours in Case B using 3 GHz processor. In the Bose gas, the speed of sound ($c_s$) is defined as $c_s:=\sqrt{\text{Li}_{\frac{5}{2}}\left(\mathfrak{Z}\right)/\text{Li}_{\frac{3}{2}}\left(\mathfrak{Z}\right)\gamma RT}$, where $\gamma=5/3$. As a result, Mach number in the uniform flow is $M_\infty=2.34$ in Case A and $M_\infty=2.47$ in Case B, respectively. Additionally, the VHS molecule with $\xi=5/9$ is calculated. Knudsen number in the uniform flow ($\text{Kn}_\infty$) is set as $\text{Kn}_\infty=2.65 \times 10^{-2}$ in Case A and $\text{Kn}_\infty=9.51 \times 10^{-2}$ in Case B, because $\text{Kn}_\infty|_{\mathfrak{Z}\rightarrow 0}=0.12$ is set in Cases A and B.}\\
\textcolor{black}{The schematic of the flow-field is shown in Fig. 4. The lattices are set as $i \in [0,121)$ and $j \in [0,60)$, in which $i \in \mathbb{Z}_+$ corresponds to the circumferential direction and $j \in \mathbb{Z}_+$ correspond to the radial direction. Thus, $j=0$ corresponds to grids on the wall. In particular. we focus on profiles of macroscopic quantities along the stagnation streamline (SSL), which corresponds to $\tilde{Y}=0 \wedge \tilde{X} \le -1$, as shown in Fig. 4. The fugacity once decreases backward the shock wave and increases toward the wall, as shown in the contour of $\mathfrak{Z}$ in Fig. 4. Consequently, the quantum effects are once weaken backward the shock wave. The total number of sample particles in the calculation domain is about $1.2\times 10^6$ in Cases A and B. The primary reason for longer CPU time in Case A than that in Case B is caused by the fact that $\int_{\bm{x} \in \mathbb{X}^3} \nu_{\max} d \bm{x}$ in Case A ($\mathbb{X}^3$: calculation domain) is larger than that in Case B owing to Eq. (3), exclusively.}\\
\textcolor{black}{The completely diffusive wall reflects molecules, which collide with the wall, in accordance with the B.E. distribution, whose temperature and fugacity are set as $\tilde{T}_w$ and $\mathfrak{Z}_w$. $\tilde{T}_w$ is calculated using $\tilde{\rho}_w$ and $\mathfrak{Z}_w$ from Eq. (9). Here, $\tilde{\rho}_w$ is calculated by}
\begin{eqnarray*}
\tilde{\rho}_w\left(\tilde{t}+\Delta \tilde{t}, i,0\right):=\left|\bigcup_{\bm{x}_\ell \left(\tilde{t}+\Delta \tilde{t}_1\right) \in  A_{i,0} \cap v_\ell^{\perp}\left(\tilde{t}+\Delta \tilde{t}_1\right)<0 } \ell\right|/\left(N_s V_{i,0}\right),~~~\left(0<\Delta \tilde{t}_1 \le \Delta \tilde{t},~\ell\in \mathbb{Z}_+\right),
\end{eqnarray*}
\textcolor{black}{where $\ell$ corresponds to the $\ell$-th sample particles in the calculation domain, $\left(i,0\right)$ corresponds to the address of the lattice, which is adjacent to the wall, $V_{i,0}$ is the volume of the lattice with the address $(i,0)$, $A_{i,0}$ is the surface area of the wall, which is included in a lattice $\left(i,0\right)$, $v_\ell^\perp$ is the molecular velocity, which is decomposed to the normal direction to the wall and $N_s$ is the number of sample particles, which corresponds to the number density in the uniform flow}.\\   
\textcolor{black}{Figure 5 shows profiles of $\tilde{\rho}$ and $\tilde{T}$ ($y_1$ axis), $\tilde{u}$ ($y_2$ axis) and $\tilde{\mathfrak{Z}}:=\mathfrak{Z}_\infty/\mathfrak{Z}$ ($y_3$ axis) along the SSL in Case A. $\mathfrak{Z}$ decreases inside the shock wave and its minimum value at point (B) ($-\tilde{X}=2.06$), whereas $\mathfrak{Z}$ increases behind the shock wave in the range of $1<-\tilde{X}<2.06$. The marked decrease of $\mathfrak{Z}$ in the thermal boundary layer ($1<-\tilde{X} \le 1.5$) is caused by the marked increase of the density owing to Eq. (8). Figure 5 shows that the completely diffusive wall works as the heating wall, because $\tilde{T}$ increases and $\tilde{u}=-0.06$ at point (A) ($-\tilde{X}=1.02$).}\\
\textcolor{black}{Figure 6 shows $\tilde{f}\left(\tilde{v}_x\right)$ versus $\tilde{v}_x$ at points (A)-(D) on the SSL together with $\tilde{f}^{\text{\tiny{eq}}}\left(\tilde{v}_x\right)$ and $\tilde{f}_{\text{MB}}\left(\tilde{v}_x\right)$, where locations of points (A)-(D) in Case A are shown in Fig. 5. $\tilde{f}\left(\tilde{v}_x\right)=\tilde{f}^{\text{\tiny{eq}}}\left(\tilde{v}_x\right)$ is obtained in the uniform flow, as shown at point (D) in Case A. $\tilde{f}\left(\tilde{v}_x\right)$ deviates from $\tilde{f}^{\text{\tiny{eq}}}\left(\tilde{v}_x\right)$ at point (C), which corresponds to the forward regime of the shock wave. As observed in the profile of $\tilde{f}\left(\tilde{v}_x\right)$ for the classical gas \cite{Yano3}, $\tilde{f}\left(\tilde{v}_x\right) \gg \tilde{f}^{\text{\tiny{eq}}}\left(\tilde{v}_x\right)$ is obtained at the negative velocity in the forward regime of the shock wave. Additionally, $\tilde{f}\left(\tilde{v}_x\right) \simeq \tilde{f}^{\text{\tiny{eq}}}\left(\tilde{v}_x\right)$ is obtained at point (B), which corresponds to the backward regime of the shock wave and location of the peak value of $\tilde{\mathfrak{Z}}$. Similarly, $\tilde{f}\left(\tilde{v}_x\right) \simeq \tilde{f}^{\text{\tiny{eq}}}\left(\tilde{v}_x\right)$ is obtained at point (A), which corresponds to the lattice on the wall. As a result, the discontinuity of the distribution function at both sides of $\tilde{v}_x=0$ on the wall \cite{Cercignani} is dismissed by the rapid equilibration inside the lattice, which is adjacent to the wall, owing to the small Kn as a result of the decrement of the fugacity in the vicinity of the wall. At points (A)-(D), $\tilde{f}\left(\tilde{v}_x\right)$ is clearly different from $\tilde{f}_{\text{MB}}\left(\tilde{v}_x\right)$.}\\
\textcolor{black}{Finally, we investigate the effect of $\mathfrak{Z}_\infty$ on profiles of $\tilde{\rho}$, $\tilde{u}$ and $\tilde{T}$ along the SSL by comparing profiles of $\tilde{\rho}$, $\tilde{u}$ and $\tilde{T}$ along the SSL in Case A with those in Case B. The left frame of Fig. 7 shows profiles of $\tilde{\rho}$ ($y_1$ axis) and $\tilde{T}$ ($y_2$ axis) along the SSL in Case B (solid lines) together with those in Case A (dashed lines), whereas the right frame of Fig. 7 shows the profile of $\tilde{\mathfrak{Z}}$ along the SSL in Case B (solid line) together with that in Case A (dashed line). The location of the shock wave in Case B is more distant from the wall than that in Case B. $\tilde{T}$ behind the shock wave ($-\tilde{X}=2.23$) in Case B is lower than that in Case A. Such a tendency conflicts with our conjecture that the temperature behind the shock wave in Case B is higher than that in Case A owing to $\left(M_\infty\right)_{\text{Case A}}<\left(M_\infty\right)_{\text{Case B}}$ and $\left(C_V\right)_{\text{Case A}}>\left(C_V\right)_{\text{Case B}}$. Such a conflict of our conjecture is described by the fact that $\tilde{T}$ at point (A) in Case B is much smaller than that in Case A. In short, the heating rate via the completely diffusive wall in Case B is smaller than that in Case A. Consequently, $\tilde{T}$ behind the shock wave ($-\tilde{X}=2.23$) in Case B is lower than that in Case A owing to the shock-thermal boundary interaction. The thickness of the shock wave in Case B is similar to that in Case, although $\left(\text{Kn}_\infty\right)_{\text{Case A}}<\left(\text{Kn}_\infty\right)_{\text{Case B}}$. Such a similarity is described by the fact that the increases of $\mathfrak{Z}$ inside the shock wave contributes to the decrease of Kn from Eq. (5) in Case A. The maximum value of $\tilde{\mathfrak{Z}}$ backward the shock wave ($-\tilde{X}=2.4$) in Case B is larger that at point (B) in Case A, whereas the minimum value of $\tilde{\mathfrak{Z}}$ in the vicinity of the wall ($-\tilde{X}=1.02$) in Case B is smaller than that at point (A) in Case A.}
\begin{figure}
\includegraphics[width=0.75\linewidth]{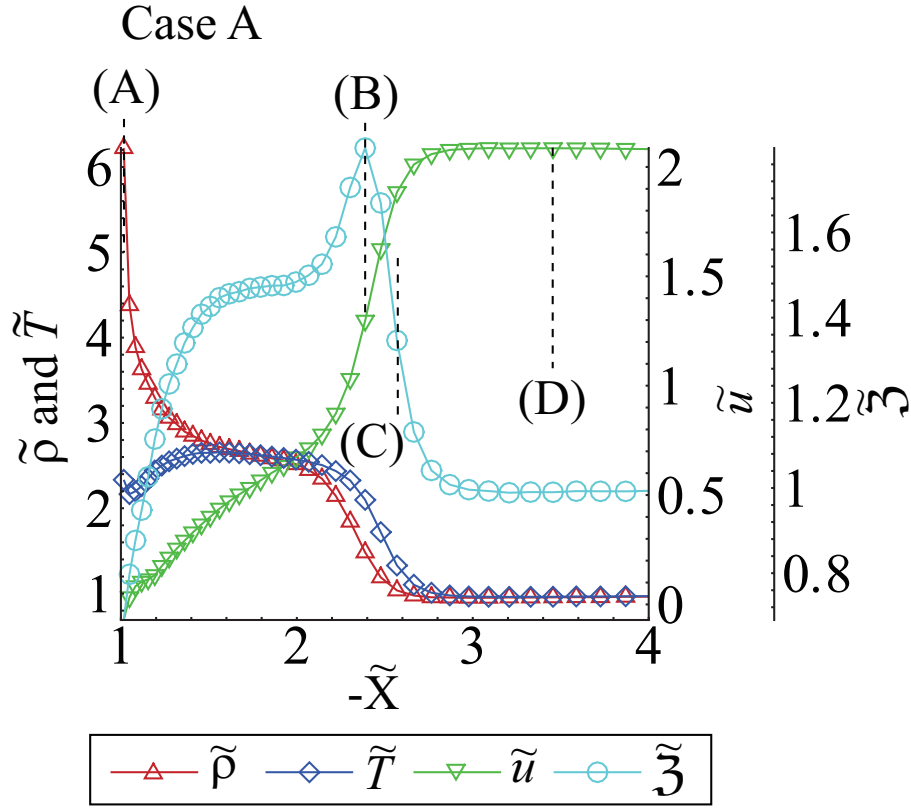}\\
\caption{Profiles of $\tilde{\rho}$ and $\tilde{T}$ ($y_1$ axis), $\tilde{u}$ ($y_2$ axis) and $\tilde{\mathfrak{Z}}:=\mathfrak{Z}_\infty/\mathfrak{Z}$ ($y_3$ axis) along the SSL in Case A.}
\end{figure}
\begin{figure}
\includegraphics[width=1.0\linewidth]{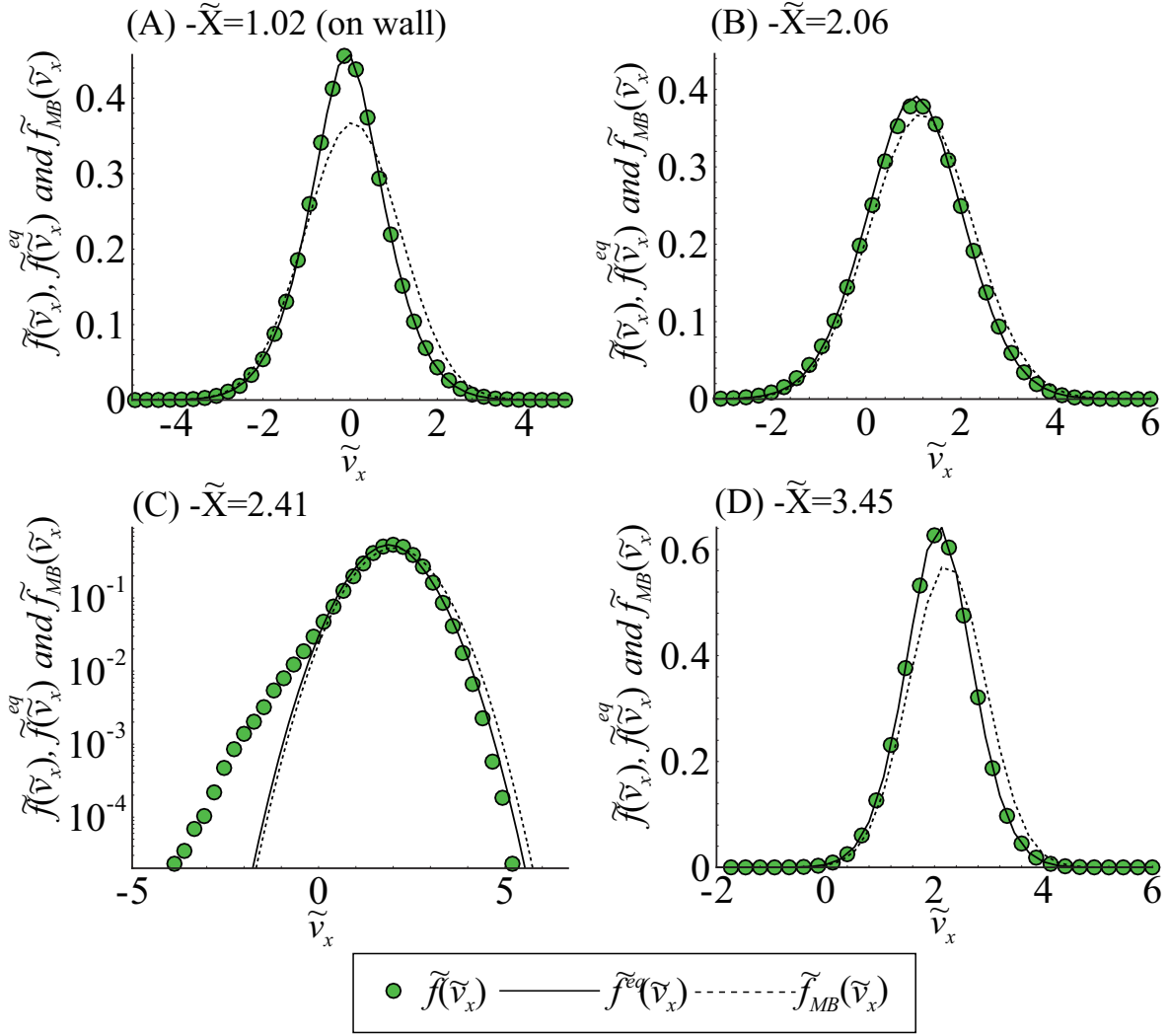}\\
\caption{$\tilde{f}\left(\tilde{v}_x\right)$ versus $\tilde{v}_x$ at points (A)-(D) on the SSL together with $\tilde{f}^{\text{eq}}\left(\tilde{v}_x\right)$ and $\tilde{f}_{\text{MB}}\left(\tilde{v}_x\right)$}
\end{figure}
\begin{figure}
\includegraphics[width=1.0\linewidth]{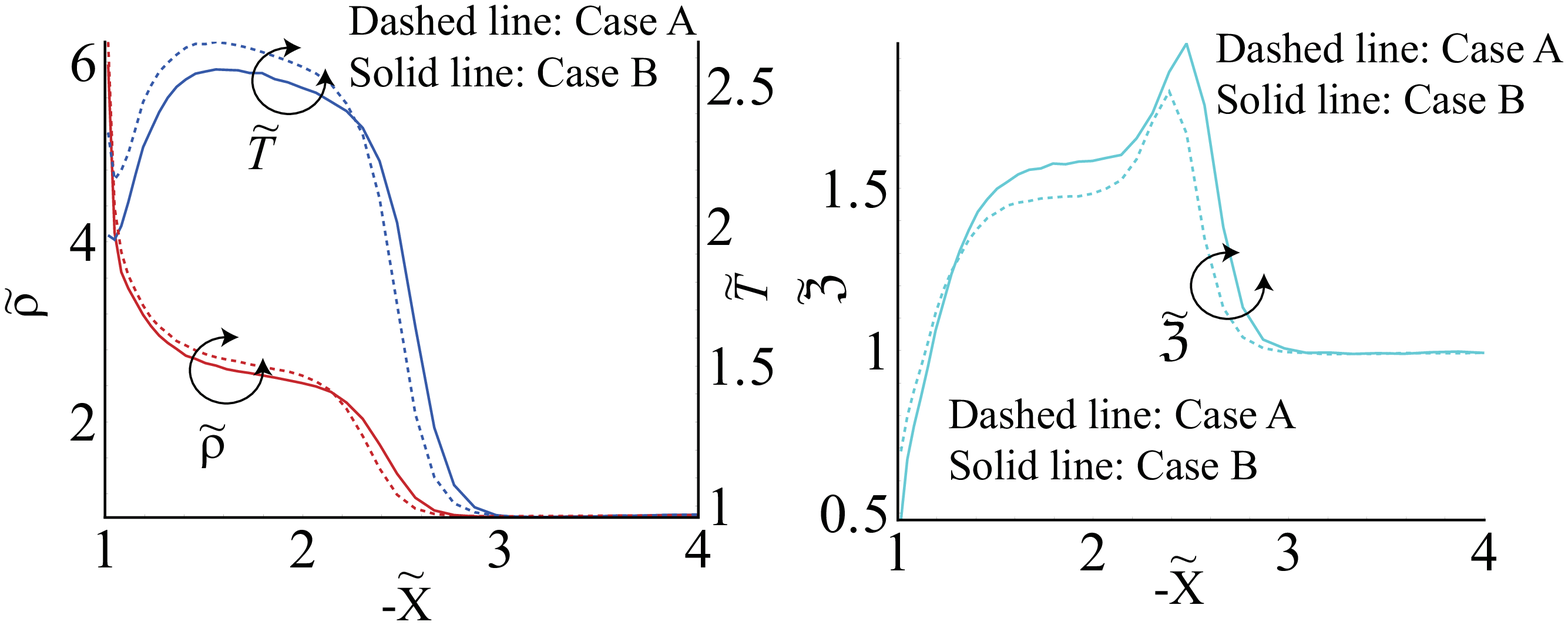}\\
\caption{Profiles of $\tilde{\rho}$ ($y_1$ axis) and $\tilde{T}$ ($y_2$ axis) along the SSL in Case B (solid lines) together with those in Case A (dashed lines) in the left frame, and profile of $\tilde{\mathfrak{Z}}$ along the SSL in Case B (solid line) together with that in Case A (dashed line) in the right frame.}
\end{figure}
\section{Conclusions}
In this paper, we proposed new numerical method to calculate the dilute and non-condensed quantum gas \textcolor{black}{in practical time}, accurately, using the U-U model equation on the basis of the DSMC method. Numerical results certainly showed that our numerical method surely enables us to calculate the dilute quantum gas dynamics in practical time and obtain the accurate thermalization using a small number of sample particles. The viscosity coefficient for the U-U model equation, which is calculated using the DSMC method on the basis of Green-Kubo expression, is similar to that for the U-U equation in the previous study in the small fugacity regime. \textcolor{black}{Additionally, the bulk relation between the viscosity coefficient and fugacity, which is obtained using the U-U model equation, is similar to that obtained using the U-U equation by Nikuni-Griffin \cite{Nikuni}. Finally, we calculated the shock layer of the dilute Bose gas using the U-U model equation in practical time. The quantum effect is weaken backward the shock wave owing to the decrease of the fugacity, whereas the quantum effect is strengthen toward the wall owing to the increase of the fugacity. Such an increase of the fugacity toward the wall is caused by the marked increase of the density toward the wall, because the fugacity increases in accordance with the increase of the density}. As a result, our numerical method enables us to calculate the strongly nonequilibrium flow. The author concludes that our numerical method must be applied to the numerical analysis of the dilute and non-condensed quantum gas, aggressively, when the dilute and non-condensed quantum gas is expected to be calculated using the DSMC method \cite{Jackson} \cite{Cerboneschi} \cite{Wade} \cite{Toschi} \cite{Urban} \cite{Goulko}, because the calculation of the accurate thermalization of the dilute quantum gas by solving the U-U equation on the basis of the DSMC method will be difficult even with the most advanced parallel computers and the incorrect thermalization with a small number of sample particles, which are used to solve the U-U equation, degrades the \textcolor{black}{accurate} estimation of quantum effects in the dilute quantum gas. As with the reproduction of accurate transport coefficients via the U-U model equation, further modification of the U-U model equation is required on the basis of Grad's 13 moment expansion of the distribution function, as discussed in appendix B.
\begin{acknowledgments}
The author acknowledges the emeritus professor Dr. Leif Arkeryd (Mathematical Science, Chalmers University) for useful comments.
\end{acknowledgments}
\begin{appendix}
\section{Comment on H theorem}
\textcolor{black}{Once the U-U equation is modified with the U-U model equation, we must reconsider H theorem for the U-U model equation. Needless to say, the definition of the entropy for the quantum gas is written as $S_{\text{\tiny{UU}}}:=\int_{\mathbb{V}^3} f\left(\bm{v}\right) \ln f\left(\bm{v}\right)\left[1-\theta f\left(\bm{v}\right)\right]^{-1} d \bm{v}$. Such a definition  of the entropy enables us to prove H theorem from the symmetry relation of the collisional term in the U-U equation such as}
\begin{eqnarray}
&&D_t S_{\text{\tiny{UU}}}:=\int_{\mathbb{V}^3} \left[\partial_t f\left(\bm{v}\right)+\bm{v} \cdot \bm{\nabla}f \left(\bm{v}\right)\right] \ln \left(\frac{f\left(\bm{v}\right)}{1-\theta f\left(\bm{v}\right)}\right) d \bm{v} \nonumber \\
&=&\frac{1}{4} \int_{I_\chi \times I_\epsilon} \int_{\mathbb{V}_1^3 \times \mathbb{V}^3} \left[f\left(\bm{v}^\prime\right)f\left(\bm{v}_1^\prime\right)\left(1-\theta f\left(\bm{v}\right)\right)\left(1-\theta f\left(\bm{v}_1\right)\right) \right. \nonumber \\
&&\left. -f\left(\bm{v}\right)f\left(\bm{v}_1\right)\left(1-\theta f\left(\bm{v}^\prime\right)\right)\left(1-\theta f\left(\bm{v}^\prime_1\right)\right)\right] \nonumber \\
&&\times \ln \left[\frac{f\left(\bm{v}\right)f\left(\bm{v}_1\right)\left(1-\theta f\left(\bm{v}^\prime\right)\right)\left(1-\theta f\left(\bm{v}_1^\prime\right)\right)}{f\left(\bm{v}^\prime\right)f\left(\bm{v}_1^\prime\right)\left(1-\theta f\left(\bm{v}\right)\right)\left(1-\theta f\left(\bm{v}_1\right)\right)}\right] g \sigma \sin \chi d \epsilon d\chi d \bm{v}_1  d \bm{v} \le 0,
\end{eqnarray}
\textcolor{black}{The definition of the entropy for the quantum gas, namely, $S_{\text{\tiny{UU}}}$ does not hold for the U-U model equation, because the collisional term of the U-U equation is modified using the thermally equilibrium distribution function. Then, we define the entropy for the U-U model equation such as:}
\begin{eqnarray}
S_{\text{\tiny{UUM}}}:=\int_{\mathbb{V}^3} f\left(\bm{v}\right) \ln \frac{f\left(\bm{v}\right)}{1-\theta f^{\text{eq}}\left(\bm{v}\right)} d \bm{v}
\end{eqnarray}
\textcolor{black}{In a similar way to Eq. (A1), we prove H theorem for the U-U model equation such as}
\begin{eqnarray}
&&D_t S_{\text{\tiny{UUM}}}:=\int_{\mathbb{V}^3} \left[\partial_t f\left(\bm{v}\right)+\bm{v} \cdot \bm{\nabla}f \left(\bm{v}\right)\right] \ln \left(\frac{f\left(\bm{v}\right)}{1-\theta f^{\text{eq}}\left(\bm{v}\right)}\right) d \bm{v} \nonumber \\
&=&\frac{1}{4} \int_{I_\chi \times I_\epsilon} \int_{\mathbb{V}_1^3 \times \mathbb{V}^3} \left[f\left(\bm{v}^\prime\right)f\left(\bm{v}_1^\prime\right)\left(1-\theta f^{\text{eq}}\left(\bm{v}\right)\right)\left(1-\theta f^{\text{eq}}\left(\bm{v}_1\right)\right) \right. \nonumber \\
&&\left. -f\left(\bm{v}\right)f\left(\bm{v}_1\right)\left(1-\theta f^{\text{eq}}\left(\bm{v}^\prime\right)\right)\left(1-\theta f^{\text{eq}}\left(\bm{v}^\prime_1\right)\right)\right] \nonumber \\
&&\times \ln \left[\frac{f\left(\bm{v}\right)f\left(\bm{v}_1\right)\left(1-\theta f^{\text{eq}}\left(\bm{v}^\prime\right)\right)\left(1-\theta f^{\text{eq}}\left(\bm{v}_1^\prime\right)\right)}{f\left(\bm{v}^\prime\right)f\left(\bm{v}_1^\prime\right)\left(1-\theta f^{\text{eq}}\left(\bm{v}\right)\right)\left(1-\theta f^{\text{eq}}\left(\bm{v}_1\right)\right)}\right] g \sigma \sin \chi d \epsilon d\chi d \bm{v}_1 d \bm{v}  \le 0,
\end{eqnarray}
\textcolor{black}{In Eqs. (A1) and (A3), $D_t S_{\text{\tiny{UU}}}=D_t S_{\text{\tiny{UUM}}}=0$ is realized, when $f\left(\bm{v}\right)=f^{\text{eq}}\left(\bm{v}\right)$. From above discussion, the directivity of $f\left(\bm{v}\right)$ toward $f^{\text{eq}}\left(\bm{v}\right)$ via binary collisions is confirmed in the U-U model equation. We, however, remind that proofs of the negativity of left hand sides of Eqs. (A1) and (A3), ($D_t S_{\text{\tiny{UU}}} \le 0$ and $D_t S_{\text{\tiny{UUM}}} \le 0$) are still mathematically open problem \cite{Villani}.}
\section{Comments on improvement of U-U model equation}
\textcolor{black}{The dissipation process of $f\left(\bm{v}\right)$ toward to $f^{\text{eq}}\left(\bm{v}\right)$ in the U-U equation is clearly different from that in the U-U model equation, so that the transport coefficients such as the viscosity coefficient and thermal conductivity for the quantum gas in the U-U equation are presumably different from those in the U-U model equation. Grad's 13 moment equations, which are derived from the U-U equation, are obtained by substituting $f\left(\bm{v}\right)=f_{13}\left(\bm{v}\right)$, $\bm{\Xi}\left(\bm{v}\right):=\left(\hat{h}^3/m\right)\left(1,\bm{v},C^2/3,C_i C_j-\delta_{ij}C^2/3,\bm{C} C^2/2\right)$ and $\bm{a}_{13}:=\int_{\mathbb{V}^3} f\left(\bm{v}\right) \bm{\Xi}\left(\bm{v}\right) d \bm{v}=\int_{\mathbb{V}^3} f_{13}\left(\bm{v}\right) \bm{\Xi}\left(\bm{v}\right) d \bm{v}=\left(\rho,\rho \bm{u},p,p_{ij},q_i\right)$ ($p_{ij}$: pressure deviator, $q_i$: heat flux) into Eq. (1) such as}
\begin{eqnarray}
D_t \bm{a}_{13}:&=&\int_{\mathbb{V}^3} \left[\partial_t f_{13}\left(\bm{v}\right)+\bm{v} \cdot \bm{\nabla}f_{13} \left(\bm{v}\right)\right] \bm{\Xi}\left(\bm{v}\right) d \bm{v} \nonumber \\
&=&\frac{1}{2} \int_{I_\chi \times I_\epsilon} \int_{\mathbb{V}_1^3 \times \mathbb{V}^3} f_{13}\left(\bm{v}\right)f_{13}\left(\bm{v}_1\right)\underline{\left(1-\theta f_{13}\left(\bm{v}^\prime\right)\right)\left(1-\theta f_{13}\left(\bm{v}^\prime_1\right)\right)} \nonumber \\
&&\times \left[\bm{\Xi}\left(\bm{v}^\prime\right)+\bm{\Xi}\left(\bm{v}_1^\prime\right)-\bm{\Xi}\left(\bm{v}\right)-\bm{\Xi}\left(\bm{v}_1\right)\right]
 g \sigma \sin \chi d \epsilon d\chi d \bm{v}_1 d \bm{v},
\end{eqnarray}
\textcolor{black}{where $f_{13}\left(\bm{v}\right)$ was calculated by the author \cite{Yano1} as}
\begin{eqnarray}
f_{13}\left(\bm{v}\right)&=&f^{\text{eq}}\left(\bm{v}\right)+f^{\text{eq}}\left(\bm{v}\right)\left(1-\theta f^{\text{eq}}\left(\bm{v}\right)\right) \nonumber \\
&& \times \left[\frac{p_{ij}}{2p} \left(\tilde{C}_i \tilde{C}_j-\frac{1}{3} \delta_{ij}\tilde{C}^2\right)+\frac{q_i C_i}{5}\frac{\mathfrak{B}^{-1}}{pRT}\left(\tilde{C}^2-5\frac{\text{Li}_{\frac{5}{2}}}{\text{Li}_{\frac{3}{2}}}\right)\right],
\end{eqnarray}
\textcolor{black}{where $\mathfrak{B}:=(7/2)\text{Li}_{7/2}-(5/2)\text{Li}_{5/2}$, in which $\text{Li}_k:=\text{Li}_k\left(-\mathfrak{Z}\theta\right)$ is the polylogarithm.\\
Similarly, Grad's 13 moment equations, which are derived from the U-U model equation, are obtained as}
\begin{eqnarray}
D_t \bm{a}_{13}:&=&\int_{\mathbb{V}^3} \left[\partial_t f_{13}\left(\bm{v}\right)+\bm{v} \cdot \bm{\nabla}f_{13} \left(\bm{v}\right)\right] \bm{\Xi}\left(\bm{v}\right) d \bm{v} \nonumber \\
&=&\frac{1}{2} \int_{I_\chi \times I_\epsilon} \int_{\mathbb{V}_1^3 \times \mathbb{V}^3} f_{13}\left(\bm{v}\right)f_{13}\left(\bm{v}_1\right)\underline{\left(1-\theta f^{\text{eq}}\left(\bm{v}^\prime\right)\right)\left(1-\theta f^{\text{eq}}\left(\bm{v}^\prime_1\right)\right)} \nonumber \\
&&\times \left[\bm{\Xi}\left(\bm{v}^\prime\right)+\bm{\Xi}\left(\bm{v}_1^\prime\right)-\bm{\Xi}\left(\bm{v}\right)-\bm{\Xi}\left(\bm{v}_1\right)\right]
 g \sigma \sin \chi d \epsilon d\chi d \bm{v}_1 d\bm{v},
\end{eqnarray}
\textcolor{black}{The differences between Eqs. (B1) and (B3) are indicated by terms with underlines. The convective form of $\bm{a}_{13}$, namely, $\int_{\mathbb{V}^3} \left[\partial_t f_{13}\left(\bm{v}\right)+\bm{v} \cdot \bm{\nabla}f_{13} \left(\bm{v}\right)\right] \bm{\Xi}\left(\bm{v}\right) d \bm{v}$ was calculated by the author \cite{Yano1}, whereas collisional terms in Eqs. (B1) and (B3) cannot be calculated owing to mathematical difficulties.}\\
\textcolor{black}{From Eqs. (B1) and (B3), Grad's 13 moment equations, which are derived from the U-U equation, are reproduced by rewriting the U-U model equation such as}
\begin{eqnarray}
&&\partial_t f\left(\bm{v}\right)+\bm{v} \cdot \bm{\nabla}f \left(\bm{v}\right) \nonumber \\
&=&\int_{I_\chi \times I_\epsilon} \int_{\mathbb{V}_1^3 \times \mathbb{V}^3} \left[f\left(\bm{v}^\prime\right)f\left(\bm{v}_1^\prime\right)\left(1-\theta f_{13}\left(\bm{v}\right)\right)\left(1-\theta f_{13}\left(\bm{v}_1\right)\right) \right. \nonumber \\
&&\left. -f\left(\bm{v}\right)f\left(\bm{v}_1\right)\left(1-\theta f_{13}\left(\bm{v}^\prime\right)\right)\left(1-\theta f_{13}\left(\bm{v}^\prime_1\right)\right)\right] \nonumber \\
&& g \sigma \sin \chi d \epsilon d\chi d \bm{v}_1 d \bm{v},
\end{eqnarray}
\textcolor{black}{Once Grad's 13 moment equations are reproduced by the U-U model equation in Eq. (B4), we expect that the viscosity coefficient and thermal conductivity, which are calculated by the U-U model equation in Eq. (B4), coincides with those calculated by the U-U equation as a result of the first Maxwellian iteration of Grad's 13 moment equations \cite{Yano1}. Meanwhile, we must remind that conditions of positivity, namely, $0<f_{13}\left(\bm{v}\right)$ and $0<1-\theta f_{13}\left(\bm{v}\right)$ in Eq. (B4) are not always satisfied, when we calculate $f_{13}$ in Eq. (B2) from numerical datum of $\bm{a}_{13}$. Provided that such conditions of positivity is satisfied, the U-U model equation in Eq. (B4) improves the U-U model equation in Eq. (B2), successfully. Numerical analysis of the U-U model equation in Eq. (B4) is set as our future study including the proof of H theorem. In principle, we must reproduce the U-U equation from the U-U model equation, when we approximate $f_{13}$ in Eq. (B2) to $f$ using infinite nonequilibrium moments beyond Grad's 13 moments.}
\end{appendix}

\end{document}